\documentclass[aps,twocolumn,prd,showpacs,nofootinbib]{revtex4}
\usepackage{amsmath}
\usepackage{graphicx}
\usepackage{dcolumn}
\usepackage{bm}
\usepackage{amssymb}
\usepackage{latexsym}
\usepackage{color}

\def\be{\begin{equation}}
\def\ee{\end{equation}}
\def\bea{\begin{eqnarray}}
\def\eea{\end{eqnarray}}

\begin{document}

\title{Reheating mechanism of Curvaton with Nonminimal Derivative Coupling to Gravity}

\author{Taotao Qiu$^{1,2}$}
\email{qiutt@mail.ccnu.edu.cn}

\author{Kaixi Feng$^3$}
\email{fkaixi@itp.ac.cn}


\vspace{16mm}

\affiliation{$1$ Institute of Astrophysics, Central China Normal University, Wuhan 430079, China}
\affiliation{$2$ Key Laboratory of Quark and Lepton Physics (MOE) and College of Physical Science $\&$ Technology, Central China Normal University, Wuhan 430079, P.R.China}
\affiliation{$3$ Institute of Theoretical Physics, Chinese Academy of Sciences, Beijing 100190, China}

\pacs{98.80.Cq}

\begin{abstract}
In this paper, we continue our study on the curvaton model with nonminimal derivative coupling (NDC) to Einstein gravity proposed in our previous work \cite{Feng:2013pba, Feng:2014tka}, focusing on the reheating mechanism. We found that according to whether the curvaton has dominated the background after the end of inflation, it will have two different behaviors of evolution, which should be the general property of curvaton with nonminimal couplings. This will cause two different parts of reheating, which goes on via the parametric resonance process. The reheating temperature is estimated for both cases in which reheating completes before and after curvaton domination, and the constraints are quite loose compared to that of overproduction of gravitino. Finally we investigated the evolution of curvature perturbation during reheating. We have shown both analytically and numerically that the curvature perturbation will not blow up during the resonance process. 
\end{abstract}

\maketitle

\section{introduction}
Fields in our universe may have various interactions with Einstein Gravity, which are described by the non-minimal coupling terms in the action. Aside from the simplest direct ones, there can also be derivative couplings, e.g., $G_{\mu\nu}\partial^\mu\phi\partial^\nu\phi$, with $\phi$ denoting the scalar field and $G_{\mu\nu}$ the Einstein tensor, respectively. Being proposed by Amendola et al. in 1993 \cite{Amendola:1993uh}, it is lately found that such a nonminimal derivative coupling (NDC) has nice properties of the so-called ``Horndeski theories" \cite{Deffayet:2009mn}, such as to keep the equation of motion 2nd order, therefore has been paid more and more attention in recent works, see \cite{Capozziello:1999uwa, Cartier:2001is, Daniel:2007kk, Sushkov:2009hk, Granda:2009fh, Gao:2010vr, Germani:2010gm, Chen:2010ru, Lin:2011zzd, Banijamali:2012kq, Skugoreva:2013ooa, Minamitsuji:2013ura, Myung:2015xha, Yang:2015pga, Qiu:2015aha, Cai:2016gjd} for examples. 

As an interesting application, in previous works \cite{Feng:2013pba, Feng:2014tka}, we have proposed a curvaton model with such a NDC term. This term possesses a ``self-modulation" mechanism that can phenomenologically make the power spectrum of perturbations scale-invariant independent of background evolutions. This is because that in this case the kinetic term of the field couples directly to the geometric variables such as Hubble parameter, so that it can compensate the deviation of spacetime from de-Sitter and make the field ``feel" itself in the inflationary background, even if it actually is not. In \cite{Feng:2013pba}, we calculated the background as well as both scalar and tensor perturbations generated by this model, and provided several constraints on parameters considering its transfer to the curvature perturbations. While in \cite{Feng:2014tka} we investigated the full types of non-Gaussianities up to the 3-rd order. 


As a consequent work, in this paper we consider another region, namely the reheating process of this curvaton. Reheating is a very important region in the evolution of the universe, for it explains how particles and light elements can be generated, after the dilution of the inflation. The oldest discussions of reheating can be pursued to the last decades of last century, see \cite{Starobinsky:1980te, Albrecht:1982mp, Dolgov:1982th, Abbott:1982hn} for original ideas and see \cite{Kofman:1994rk, Traschen:1990sw} where the great breakthrough of parametric resonance mechanism was proposed for efficient reheating of single field inflation. As more and more models for early universe come out, more and more reheating mechanisms are also proposed, such as geometric reheating \cite{Bassett:1997az}, curvaton reheating \cite{Feng:2002nb}, modulated reheating \cite{Dvali:2003em}, bounce reheating \cite{Cai:2011ci}, and so on. Moreover, Recently there has been some discussions on the reheating mechanism for the inflaton with such non-minimal derivative coupling (NDC) terms \cite{Sadjadi:2012zp, Ohashi:2012wf, Ghalee:2013ada, Jinno:2013fka, Ema:2015oaa, Myung:2016twf}. These works raised a couple of questions and discussions about reheating in such models, which is interesting and related to the topic of this paper. For reviews on reheating mechanisms, see \cite{Bassett:2005xm, Allahverdi:2010xz}.   


In the following, we will investigate in detail the reheating process caused by the NDC curvaton model. For simplicity, we assume that the background is still given by inflation. We will mainly focus on the following questions: \\
$\bullet$ {\it How will the curvaton evolve after the end of inflation? }\\
$\bullet$ {\it Can it give efficient mechanism for particle creation (namely, does parametric resonance exist)?}\\
$\bullet$ {\it How is the reheating temperature constrained?} \\
$\bullet$ {\it How will the curvature perturbation be affected by reheating process?} \\
and by our study, we wish to discover interesting properties of nonminimal coupling curvaton reheating, and find the difference from other kinds of reheating processes, such as that caused by inflaton itself, or other minimal coupling curvatons.
 
Our paper is organized as follows: In Sec. II we briefly review the NDC curvaton model. In Sec. III we study the reheating process of our model. In Sec. IIIA we show the background evolution both before and after curvaton domination, in Sec. IIIB we study the process of parametric resonance, and in Sec. IIIC we estimate the reheating temperature. In Sec. IV we discuss about the evolution of curvature perturbation in reheating process. Sec. V comes the conclusion.

\section{the NDC curvaton model}
Following the preceding work \cite{Feng:2013pba}, one has the action including the NDC curvaton as 
\be\label{action} 
\mathcal{S}=\int\mathrm{d}^4x\sqrt{g}\Big[\frac{R}{16\pi G}+\frac{\xi}{M^2}G_{\mu\nu}\partial^\mu\varphi\partial^\nu\varphi-V(\varphi)+{\cal L}_{bg}\Big]~,
\ee 
where $\xi$ is the dimensionless coupling constant of the NDC term. ${\cal L}_{bg}$ is the Lagrangian of the background, which drives inflation, and we don't need to specify its detailed form.
It is straightforward to write down the equation of motion for the curvaton field $\varphi$, such as
\be\label{eom}
\frac{6\xi}{M^{2}}H^{2}\ddot{\varphi}+\frac{6\xi}{M^{2}}(2\dot{H}+3H^{2})H\dot{\varphi}+V_{\varphi}=0~,
\ee
and its energy density and pressure can be expressed as
\bea\label{rho}
\rho_\varphi&=&\frac{9\xi}{M^2}H^2\dot\varphi^2+V(\varphi)~,\\
\label{p} 
P_\varphi&=&-\frac{\xi}{M^2}(3H^2\dot\varphi^2+2\dot H\dot\varphi^2+4H\dot\varphi\ddot\varphi)-V(\varphi)~,
\eea
respectively. 

The background evolution of the curvaton field with various types of potential as well as the linear perturbations during inflation has been classified and briefly analyzed in \cite{Feng:2013pba}.
As has been shown there, if the curvaton field is massless, namely $V(\varphi)=0$, exactly scale-invariant power spectrum will be obtained due to the nonminimally kinetic coupling. However, since exact scale-invariance is not favored by today's data, a non-zero potential is needed which could give a mass term to $\varphi$. \footnote{For varying Hubble parameter in non-inflationary case, a nearly-constant correction could also be obtained, with a varying mass of $\varphi$.} For the simplest choice, we choose 
\be\label{potential}
V(\varphi)=\frac{1}{2}m^2\varphi^2~
\ee
as its potential. Then the equation of motion (\ref{eom}) reduces to
\be
\ddot\varphi+(3-2\epsilon)H\dot\varphi+\left(\frac{mM}{\sqrt{6\xi}H}\right)^2\varphi=0~,
\ee
where the slow-roll parameter $\epsilon$ is defined as $\epsilon\equiv-\dot H/H^2$. Moreover, one can define an effective mass of the curvaton field,
\be
\bar{m}\equiv\frac{mM}{\sqrt{6\xi}H}~.
\ee
Moreover, according to \cite{Feng:2013pba}, we have 
\be
\frac{a^2V_{\varphi\varphi}}{2z^2}=\frac{a^2m^2}{2z^2}=\frac{\Delta_1}{(\eta_\ast-\eta)^2}~,
\ee
where $z^2\simeq (3\xi/M^2)a^2H^2$ and $\Delta_1$ describes the deviation of spectral index from pure scale-invariance, $\Delta_1\simeq 3(n_s-1)/2$. From the central value of $n_s$ given by the Planck observational constraint \cite{Ade:2015lrj}, one roughly has $\Delta_1\sim {\cal O}(0.01)$. Notice also that during the inflation $aH\simeq-(\eta_\ast-\eta)^{-1}$, this furtherly gives: 
\be\label{meff}
\bar{m}\sim 0.1H_{inf}~
\ee  
during the inflation. 

\section{reheating mechanism}
\subsection{the behavior of curvaton during oscillation}
In order to investigate the reheating mechanism of the curvaton field $\varphi$, let's assume that after the inflation ends, the background field (inflaton) decays rapidly without oscillation. A typical example is that the inflaton field is a canonical scalar field dominated by its kinetic energy, such as the so-called ``quintessential inflation" \cite{Peebles:1998qn}. Therefore one has 
\be
\rho_{bg}\propto a^{-6}(t)~,~~~w_{bg}=1~.
\ee

However, as the curvaton field has a potential (\ref{potential}), it will fall down and oscillate around its minimum, and if the energy density of curvaton decrease slower than that of inflaton, the curvaton can dominate the universe, unless it reheats before domination. First of all, one can determine the averaged equation of state during its oscillation, which will be used for later analysis. To do this, one can parametrize the scale factor $a(t)$ and Hubble parameter $H(t)$ during the rapid deflation of the universe. One can have:
\be\label{paraH}
a(t)\sim t^p~,~~H(t)=\frac{p}{t}~,~~p\equiv\frac{2}{3(1+w)}~,
\ee
and $w$ should be equal to $w_{bg}$ before curvaton domination, and $w_{\varphi}$ after that. For convenience, we define $t_{eq}$ as the time when the energy densities of inflaton and curvaton are equal. For the former case, $p=1/3$ as $w_{bg}=1$, while the latter case depends on the behavior of $\varphi$ which is the solution of the equation of motion (\ref{eom}). Considering (\ref{paraH}), Eq. (\ref{eom}) can be solved as:
\be
\varphi=\Phi(t)\times\left\{J_{\frac{3(1-p)}{4}}\left(\frac{mMt^2}{2\sqrt{6\xi}p}\right)~,~J_{\frac{-3(1-p)}{4}}\left(\frac{mMt^2}{2\sqrt{6\xi}p}\right)\right\}~,
\ee
where $\Phi(t)=\Phi_0 t^{(1-3p)/2}$ is the oscillation amplitude of $\varphi$ with the initial condition of $\Phi_0$, and $J$ is the Bessel function. Note that in reheating era where $\varphi$ begins to oscillate, one has $\bar{m}>H$, which roughly gives $\frac{mMt^2}{2\sqrt{6\xi}p}>1$. Thus the Bessel function can be mimicked by the trigonometric functions, namely
\be\label{varphisol}
\varphi\approx\Phi(t)\cos\left(\frac{mMt^2}{2\sqrt{6\xi}p}\right)~.
\ee

During rapid oscillation, the kinetic and potential terms in the energy density (\ref{rho}) are of the same order, therefore one roughly has 
\be\label{rhooft}
\rho_\varphi\sim \Phi(t)^2\propto t^{1-3p}~.
\ee
For $t<t_{eq}$, $p=1/3$, so we have $\rho_\varphi\propto t^0\sim a^0$. This means that before curvaton domination, the energy density of curvaton field will behave as a constant, giving its equation of motion to be $w_\varphi=-1$. 

Since the background energy density is decreasing, the curvaton will exceed the background (at the time $t_{eq}$) and dominate the universe. For $t>t_{eq}$, the same solution of (\ref{varphisol}) applied but with $p$ be related to $w_{\varphi}$, so one cannot determine $p$, or the behavior of $\rho_\varphi$, solely by the relation (\ref{rhooft}). However, since the curvaton dominates the universe, we have another equation, namely the Friedmann equation, 
\be
H^2=\frac{8\pi G}{3}\rho_\varphi~,
\ee
This equation, together with (\ref{paraH}), tells us that $\rho_\varphi\propto t^{-2}$. Comparing it with Eq. (\ref{rhooft}), one gets $p=1$, which gives $w_\varphi=-2/3$. \footnote{Actually there is a debate in the literature about the equation of state for the nonminimal kinetic coupling field during its oscillation, see \cite{Ghalee:2013ada, Ema:2015oaa, Myung:2016twf, Sadjadi:2012zp, Jinno:2013fka}.}

\begin{figure}[ht]
\begin{center}
\includegraphics[scale=0.5]{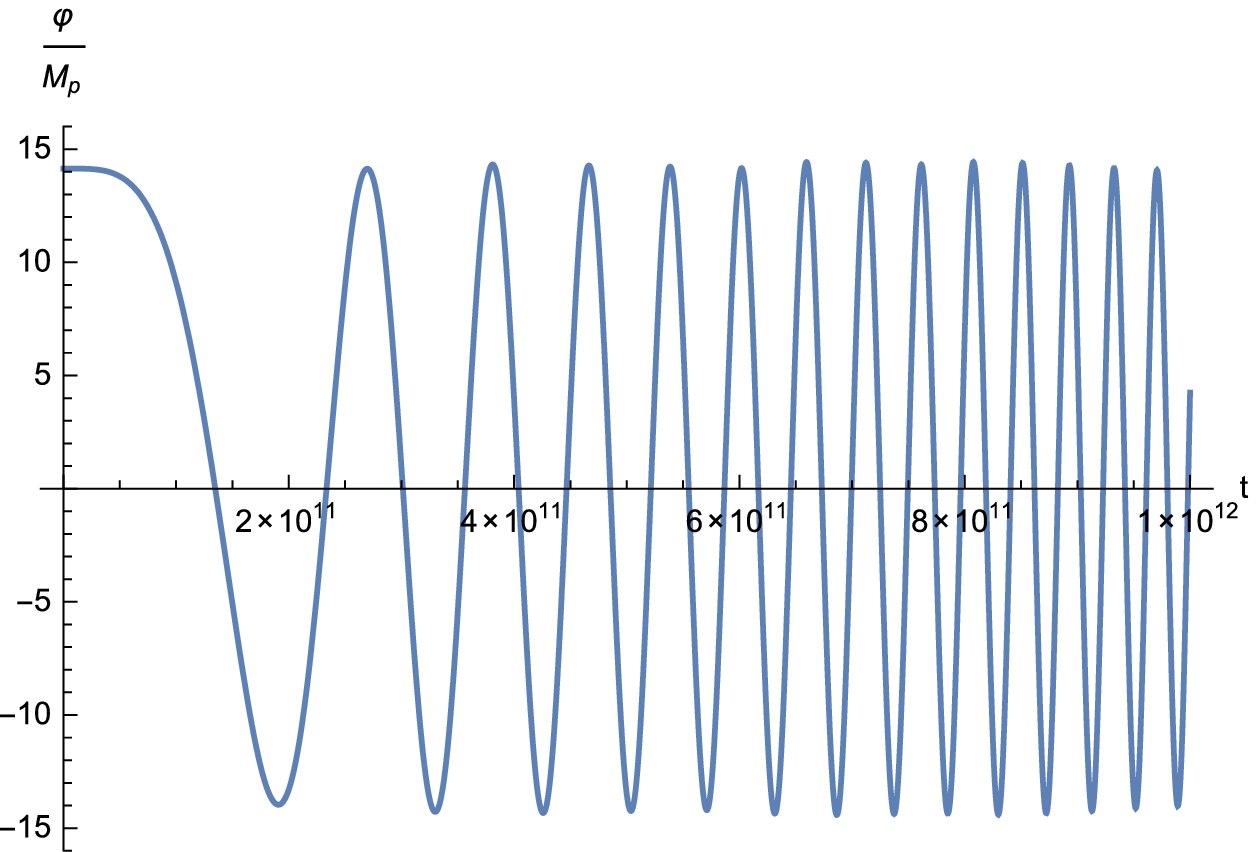}
\includegraphics[scale=0.5]{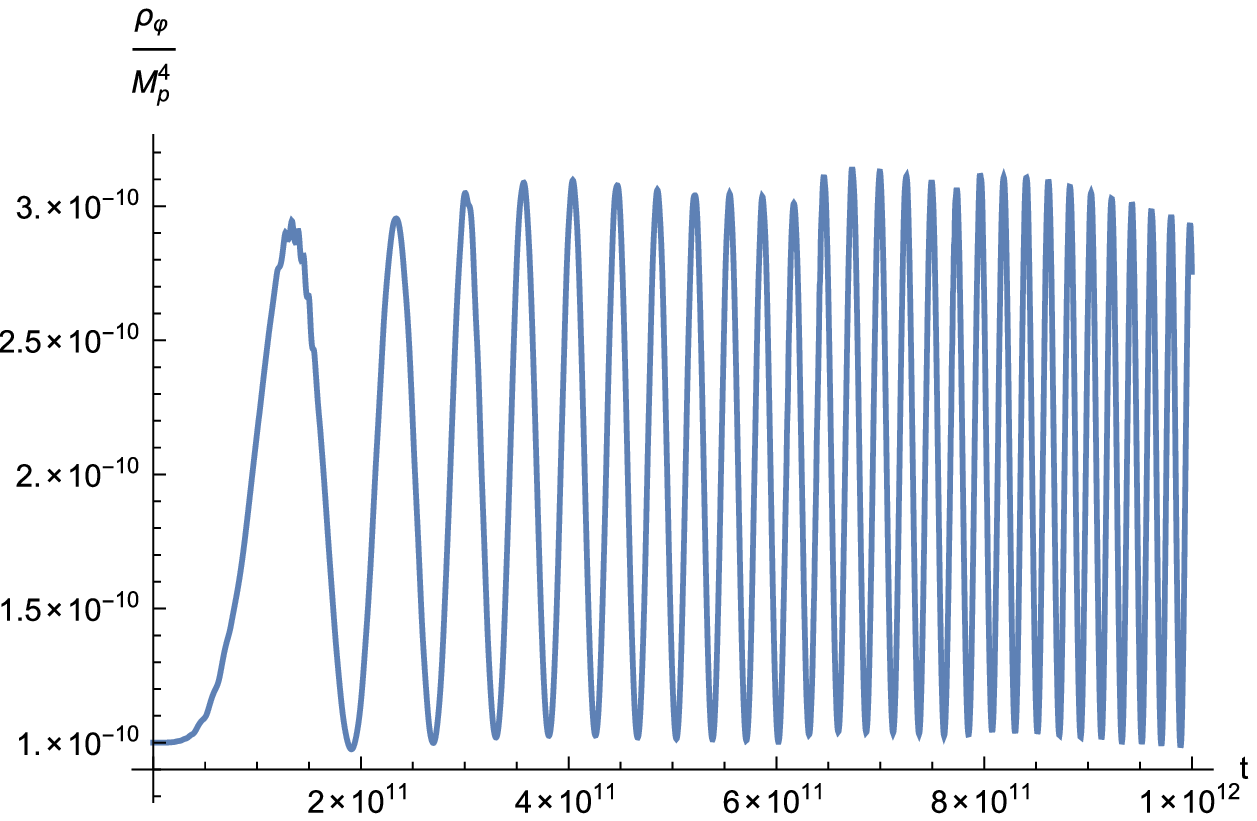}
\caption{Plots of the value of $\varphi$ and its energy density $\rho_\varphi$ in terms of cosmic time $t$. We choose the parameters as $\xi=1$, $M=10^{-4}M_p$, $m=10^{-6}M_p$. } \label{plot1}
\end{center}
\end{figure}

\begin{figure}[ht]
\begin{center}
\includegraphics[scale=0.5]{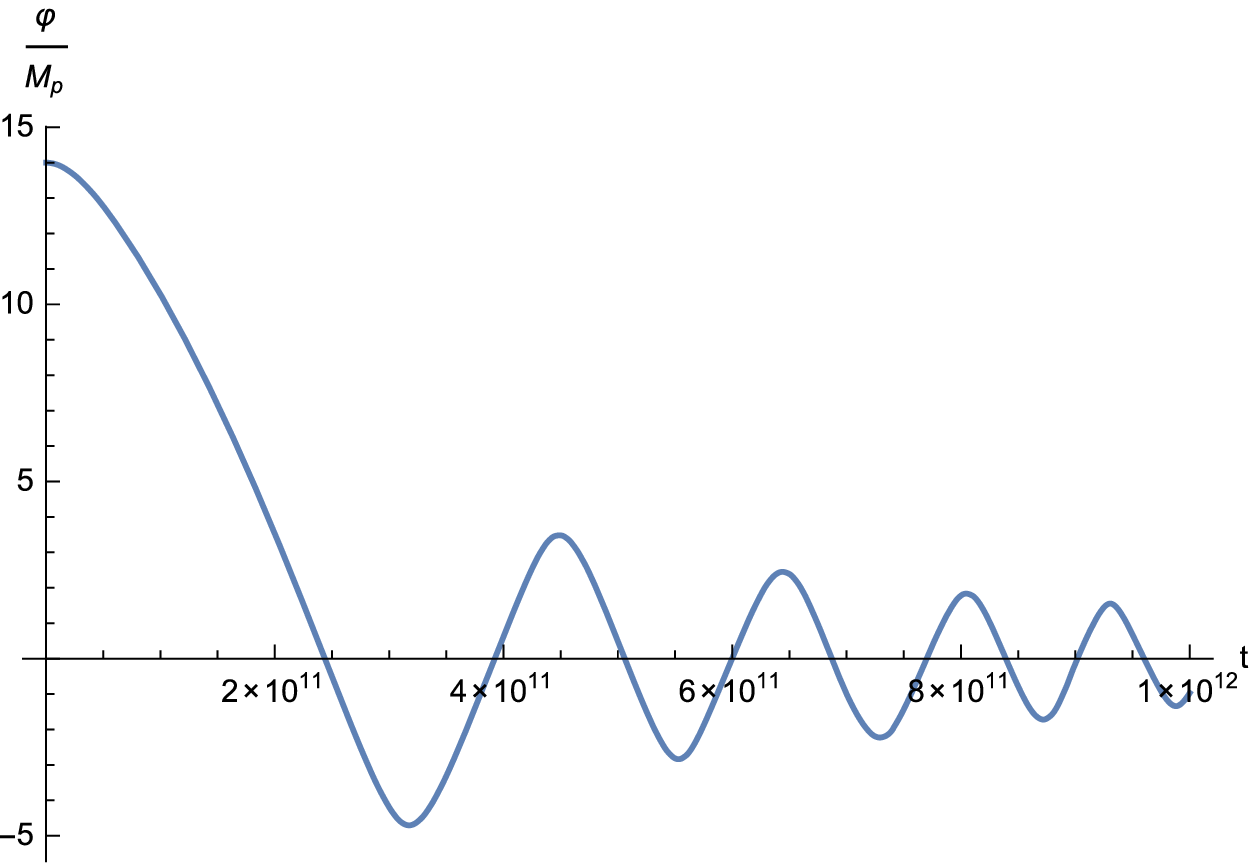}
\includegraphics[scale=0.5]{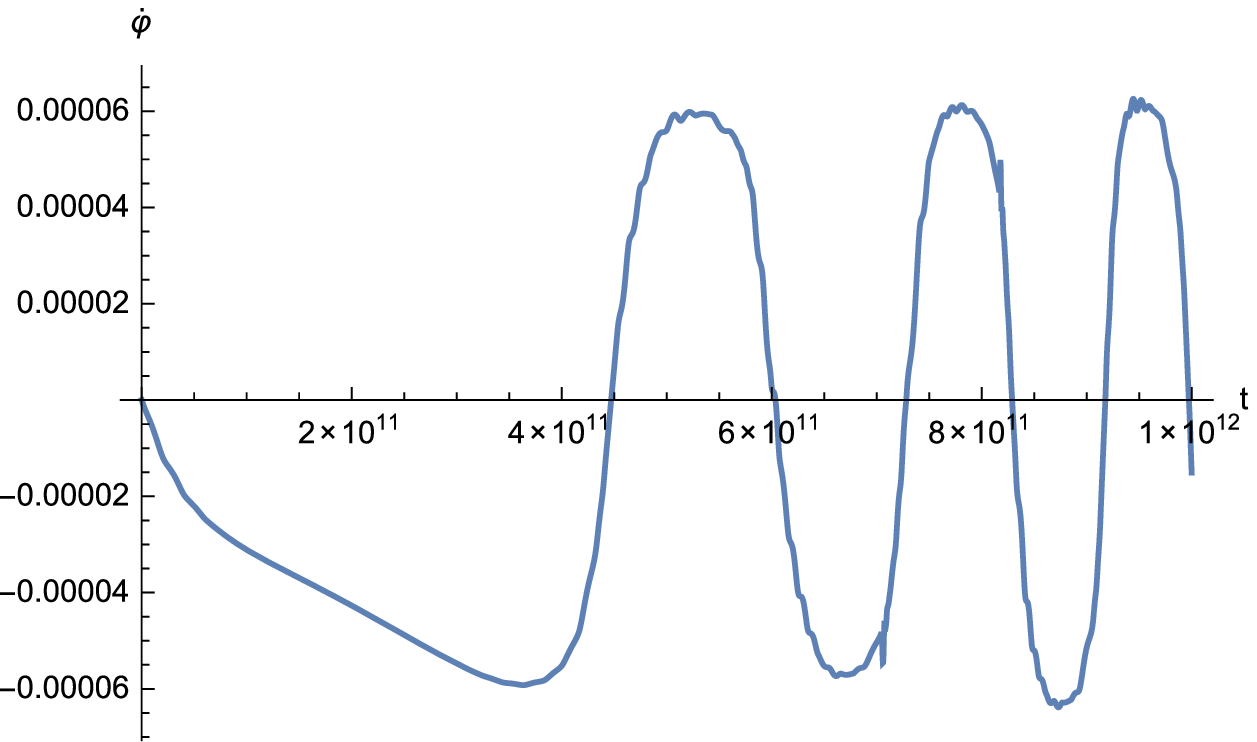}
\includegraphics[scale=0.5]{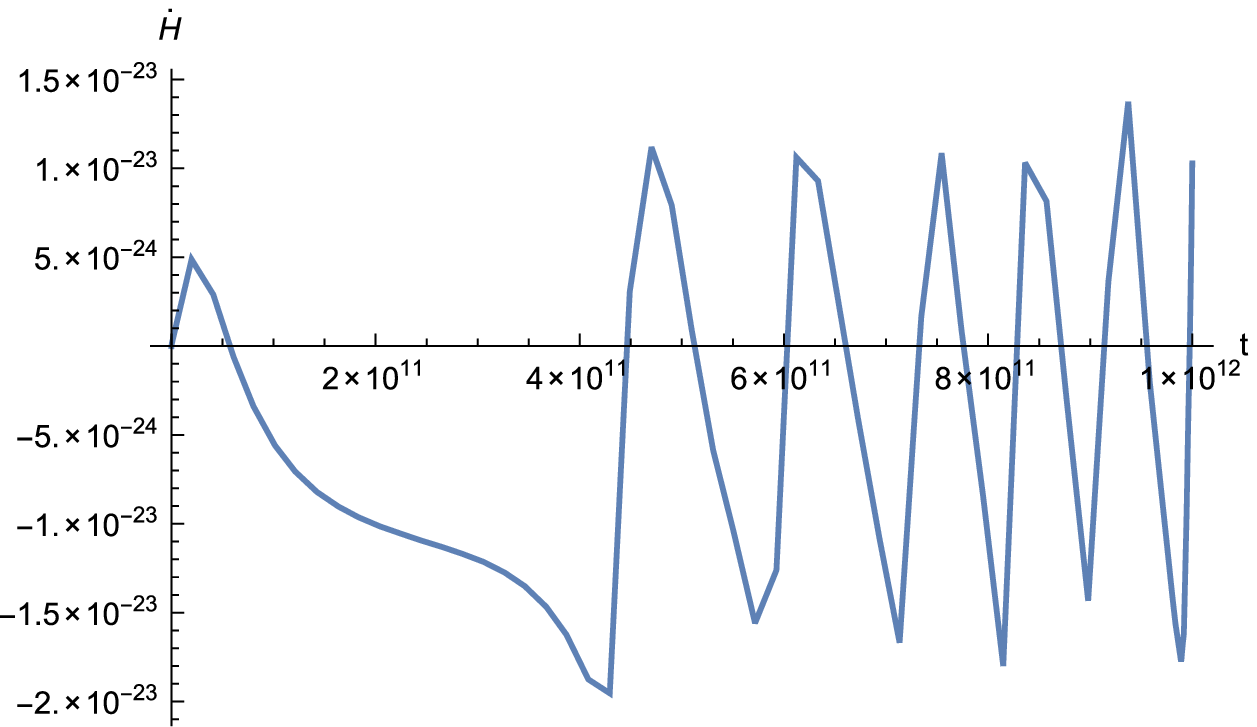}
\includegraphics[scale=0.5]{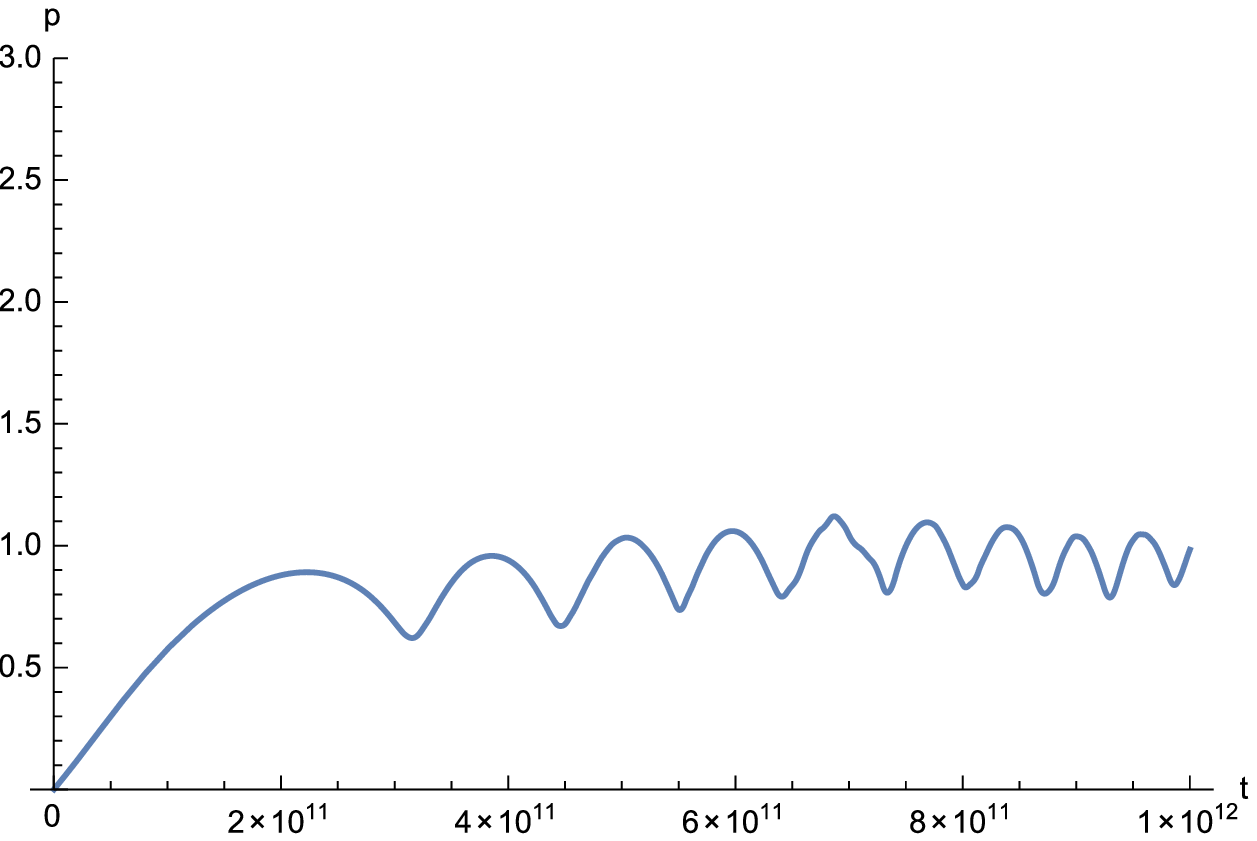}
\caption{Plots of the value of $\varphi$, $\dot\varphi$, the parameter $p$ in Eq. (\ref{paraH}) and $\dot H$ in terms of cosmic time $t$. We choose the parameters as $\xi=1$, $M=10^{-4}M_p$, $m=10^{-6}M_p$. } \label{plot2}
\end{center}
\end{figure}

In order to confirm our analysis, we performed numerical calculations for both the two cases. One can see that in Fig. \ref{plot1}, both the field value $\varphi$ and its energy density $\rho_\varphi$ oscillate with a stable amplitude, giving a constant averaged value. However, in Fig. \ref{plot2}, the field value oscillates with a damping amplitude. The Hubble parameter is therefore also oscillating, so the value of $p$ is also an oscillating function with its averaged value $\langle p\rangle=\langle H\rangle t$. From the figure it is clearly seen that $\langle p\rangle$ is at the position of about unity. We also plot velocities of $\varphi$ and $H$ for later use. 

As a side remark, we remind the reader that our result is different from that for minimal coupling curvatons \cite{Feng:2002nb}. In the latter case, if the curvaton oscillates (let's still take a mass-squared potential for curvaton as an example), the solution of curvaton will be as $\varphi=t^{-3p/2}\cos (mt)$, and $\rho_\varphi\sim t^{-3p}\sim a^{-3}$, accordingly. One can see that, no matter what value $p$ will be given, or no matter whether curvaton is dominant or not, the curvaton will behave like an ordinary matter, with $w_\varphi=0$ averagely, as long as the mass-squared potential is applied. This means the energy density will evolve synchronically with the scale factor, in terms of $t$. This result is consistent with \cite{Feng:2002nb}. However when curvaton is nonminimally coupled, this synchronicity will be violated, and the behavior before and after domination will be different. As will be seen later, this different will directly cause the different rate of particle creation. To our knowledge, this has not been given much notice in previous studies.

\subsection{parametric resonance}
In this section, we begin to discuss about how the oscillating behavior of the curvaton $\varphi$, as shown above, can be responsible for creating particles (we denote as $\chi$), namely, the preheating process. One assume the free-field Lagrangian of $\chi$ has a canonical form, and that $\varphi$ can interact with $\chi$ through some interacting term, such as $g\varphi^2\chi^2$, with a dimensionless coefficient of $g$. It gives the total form of Lagrangian of $\chi$ field as:
\be\label{lagrangianchi}
{\cal L}_\chi=-\frac{1}{2}\partial_\mu\chi\partial^\mu\chi-\frac{1}{2}m_\chi^2-\frac{1}{2}g\varphi^2\chi^2~.
\ee
It is true that one can also have the interaction term of $\chi$ and the background field (the inflaton), however here we drop this out for simplicity. Actually, as the inflaton decays quickly after inflation ends, it will have much less effects on $\chi$ than $\varphi$. Since $\chi$ is a quantum field, one can write down the Heisenberg presentation of $\chi$ as \cite{Kofman:1994rk}:
\be\label{Heisenbergchi}
\chi(t,\overrightarrow{x})=\frac{1}{(2\pi)^{3/2}}\int d^3k(\hat{a}_k\chi_k(t)e^{-i\overrightarrow{k}\cdot\overrightarrow{x}}+\hat{a}_k^\dagger\chi_k^\ast(t)e^{i\overrightarrow{k}\cdot\overrightarrow{x}})
\ee
where $\hat{a}_k$ and $\hat{a}_k^\dagger$ is the annihilation and creation operators respectively, and $k$ is the comoving wavenumber. Using this, one can also get the equation of motion for $\chi$ by simply varying the Lagrangian (\ref{lagrangianchi}):
\be\label{eomchi}
\ddot\chi_k+3H\dot\chi_k+(\frac{k^2}{a^2}+m_\chi^2+g\varphi^2)\chi_k=0~.
\ee
For the rest of this paper, we assume $m_\chi=0$ for simplicity. 

Since in our case, $\varphi(t)$ behaves as in Eq. (\ref{varphisol}), in the static limit where $a=\text{const.}$, $H=0$, Eq. (\ref{eomchi}) will reduce to the Mathieu-like equation, the solution of which is the well-known Floquet solution \cite{Kofman:1994rk}. In expanding phase such as inflation, however, one can define a new variable: $X_k(t)\equiv a^{3/2}(t)\chi_k$, and Eq. (\ref{eomchi}) can be rewritten as:
\bea\label{eomX}
&&\ddot X_k+\omega_k^2X_k=0~, \nonumber\\
&&\omega_k^2\equiv\frac{k^2}{a^2}+\frac{g\Phi_0^2}{t^{3p-1}}\cos^2\left(\frac{mMt^2}{2\sqrt{6\xi}p}\right)-\frac{3}{2}\dot H-\frac{9}{4}H^2~.
\eea
One can apply the WKB approximation to get the solution of Eq. (\ref{eomX}):
\be\label{solX}
X_k(t)=\frac{\alpha_k(t)}{\sqrt{2\omega_k}}e^{-i\int^t\omega_kdt}+\frac{\beta_k(t)}{\sqrt{2\omega_k}}e^{i\int^t\omega_kdt}~,
\ee
with $\alpha_k$ and $\beta_k$ being the time-dependent coefficient of positive and negative frequency parts, respectively. Moreover, the $\alpha_k$ and $\beta_k$ satisfy the equations:
\be\label{eomalphabeta}
\dot\alpha_k=\frac{\dot\omega_k}{2\omega_k}e^{2i\int^t\omega_kdt}\beta_k~,~~~\dot\beta_k=\frac{\dot\omega_k}{2\omega_k}e^{-2i\int^t\omega_kdt}\alpha_k~
\ee 
with additional normalization relation
\be\label{norm}
|\alpha_k|^2-|\beta_k|^2=1~.
\ee
Furthermore, the comoving occupation number of $\chi$ particles in the mode $k$ is defined by
\be
n_k\equiv\frac{\omega_k}{2}\left(\frac{|\dot X_k|^2}{\omega_k^2}+|X_k|^2\right)-\frac{1}{2}~,
\ee
Using Eq.s (\ref{solX}), (\ref{eomalphabeta}) and (\ref{norm}), one get a simple expression of $n_k=|\beta_k|^2$, and the vacuum expectation value of particle number density of $\chi$ per comoving volume is:
\be
\langle n_\chi\rangle\equiv\int_0^\infty\frac{d^3k}{(2\pi a)^3}n_k=\frac{1}{2\pi^2a^3}\int_0^\infty dk k^2|\beta_k|^2~.
\ee

It is straightforward to know how the $\chi$ particles increase with time just by integrating $|\beta_k|^2$, which is the solution of Eq. (\ref{eomalphabeta}). However, it is rather difficult, if not impossible, to have analytical solution of (\ref{eomalphabeta}) due to the complication of the dispersion relation in (\ref{solX}). In order to solve the problem, Ref. \cite{Kofman:1994rk} introduced a method, that is, to consider the solution of each period around the time point at which the source field of particle creation is zero, namely $t_j$, $j=1,2,3,...$ where $\varphi(t_j)=0$ in our case. This is because that only around $t_j$ the particle number can be dramatically changed. If we Taylor expand the function $g\varphi^2(t)$ with $\varphi$ given in (\ref{varphisol}) around $t_j$, we can get:
\be
g\varphi^2(t)=g\Phi_0^2\frac{m^2M^2t_j^{3(1-p)}}{6p^2\xi}(t-t_j)^2+{\cal O}(t-t_j)^3~.
\ee
So the leading order of $g\varphi^2(t)$ gives a parabolic potential for $\chi$, and the particle creation at $t_j$ could be viewed as the scattering of $\chi$ particles through this potential for the $j$-th time. We assume the solution of $X_k$ at the period of $t_{j-1}<t<t_j$ is
\be
X^j_k(t)=\frac{\alpha_k^j}{\sqrt{2\omega_k}}e^{-i\int^t\omega_kdt}+\frac{\beta_k^j}{\sqrt{2\omega_k}}e^{-i\int^t\omega_kdt}~,
\ee
with $\alpha_k^j$ and $\beta_k^j$ approximately constants. Following \cite{Kofman:1994rk} and noticing the relation of $n_k=|\beta|^2$, one can get a recursive relation of $n_k^j$ and $n_k^{j+1}$ as:
\be
n_k^{j+1}=n_k^je^{2\pi\mu_k^j}
\ee
with $\mu_k^j\equiv(2\pi)^{-1}\ln(1+2e^{-\pi\kappa^2}+...)$, where the ellipsis denotes the random (stochastic) terms. Here $\kappa\equiv k/(ak_\ast)$ and $k_\ast=\sqrt{\sqrt{g/(6\xi)}mMt_j^{3(1-p)/2}/p}$. In \cite{Kofman:1994rk}, $k_\ast$ is independent on $t_j$, but as is mentioned before, in our case there are two differences. One is that $p$ does not have the same value before and after curvaton domination, and the other is that the scaling of $\varphi$ on $t_j$ is different from the minimal coupling case, as was shown in (\ref{varphisol}). These causes $k_\ast\propto t_j^{1/2}$ for $t_j<t_{eq}$, while $k_\ast\propto t_j^0$ for $t>t_{eq}$. Considering also the dependence of $a$ on $t$, namely $a\propto t^{1/3}$ before $t_{eq}$ and $a\propto t$ after, one gets $\kappa\propto t^{-5/6}$ before $t_{eq}$ and $\kappa\propto t^{-1}$ after. Therefore one gets:
\bea\label{mu}
\mu_k^j&\equiv&\frac{1}{2\pi}\ln(1+2e^{-\frac{\pi}{t^{5/3}}}+...)~{\text with}~\kappa\propto t^{-5/6},\nonumber\\
\mu_k^j&\equiv&\frac{1}{2\pi}\ln(1+2e^{-\frac{\pi}{t^{2}}}+...)~{\text with}~\kappa\propto t^{-1}~.
\eea
In the limit of $\kappa^2\ll\pi^{-1}$, $\mu_k$ reduces to the trivial case in \cite{Kofman:1994rk}: $\mu_k^j\simeq(2\pi)^{-1}\ln{3}\simeq 0.18$ where we neglected the random terms.

The total number during the whole process of resonance yields
\be 
n_k=\frac{1}{2}e^{2\pi\sum_j\mu_k^j}\simeq e^{2m\int^tdt\mu_k(t)}~,
\ee
in mode $k$. In principle one can get $n_k$ by integrating $\mu_k$ according to the expressions (\ref{mu}), however unfortunately it might not be analytical, so we will refer to numerical calculations later on. The particle number density in real space is:
\bea\label{nchi}
\langle n_\chi\rangle&=&\frac{1}{2\pi^2a^3}\int_0^\infty dk k^2n_k \nonumber\\
&\sim&\frac{k_\ast^3e^{2\mu mt}}{64\pi^2a^3\sqrt{\pi\mu mt}}~,\eea
while the increasing rate of the particle number in the comoving volume is roughly estimated as:
\bea
\frac{d\left(a^3\langle n_\chi\rangle\right)}{dt}\sim\left\{
\begin{array}{l}
(\frac{3g\Phi_0^2m^2M^2}{2\xi})^{\frac{3}{4}}\frac{e^{2\mu m t}(2\mu mt+1)}{64\pi^{5/2}\sqrt{\mu m}}~{\rm for}~t<t_{eq}~,\\\\
(\frac{g\Phi_0^2m^2M^2}{6\xi})^{\frac{3}{4}}\frac{e^{2\mu m t}(4\mu mt-1)}{128\pi^{5/2}t\sqrt{\mu m t}}~{\rm for}~t>t_{eq}~.
\end{array}\right.
\eea

\begin{figure}[ht]
\begin{center}
\includegraphics[scale=0.5]{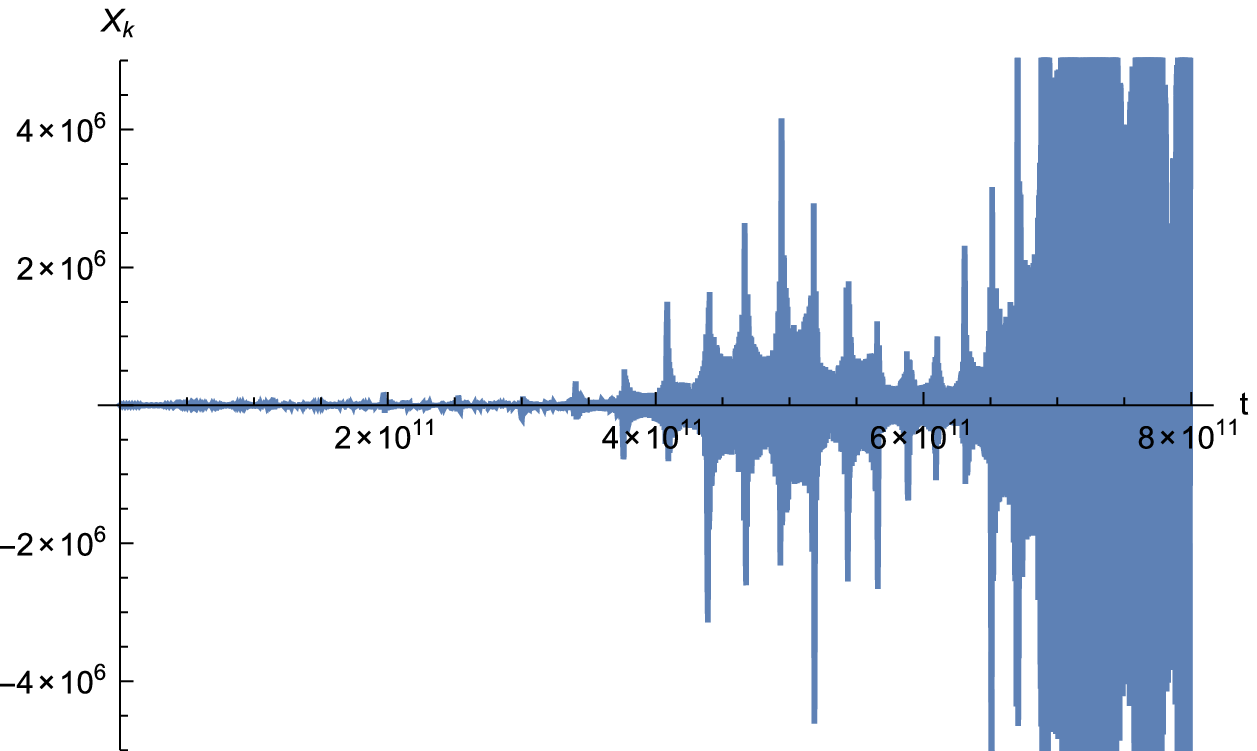}
\includegraphics[scale=0.5]{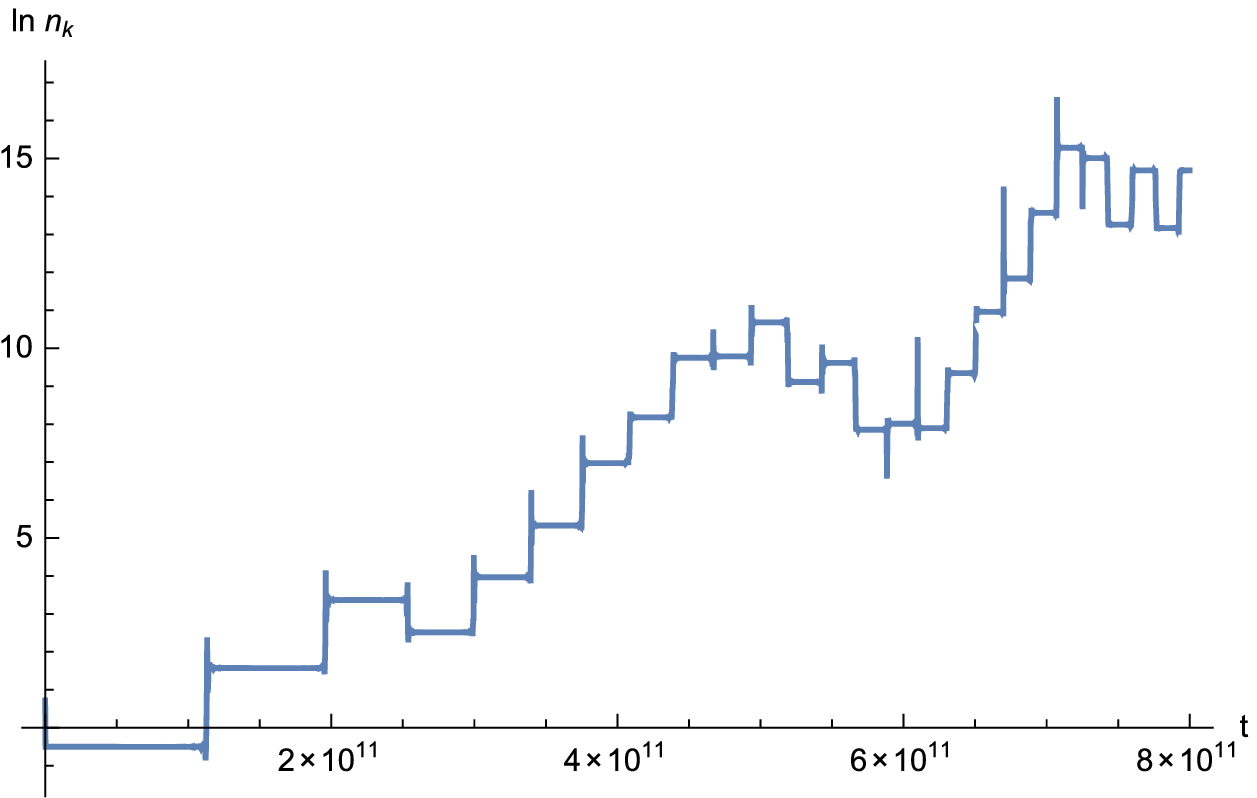}
\caption{The numerical plot of $X_k$ and $\ln n_k$ in terms of cosmic time $t$ before curvaton domination. We choose the parameters as $\xi=1$, $M=10^{-4}M_p$, $m=10^{-6}M_p$, $g=4.8\times10^{-3}$. } \label{plot3}
\end{center}
\end{figure}

\begin{figure}[ht]
\begin{center}
\includegraphics[scale=0.5]{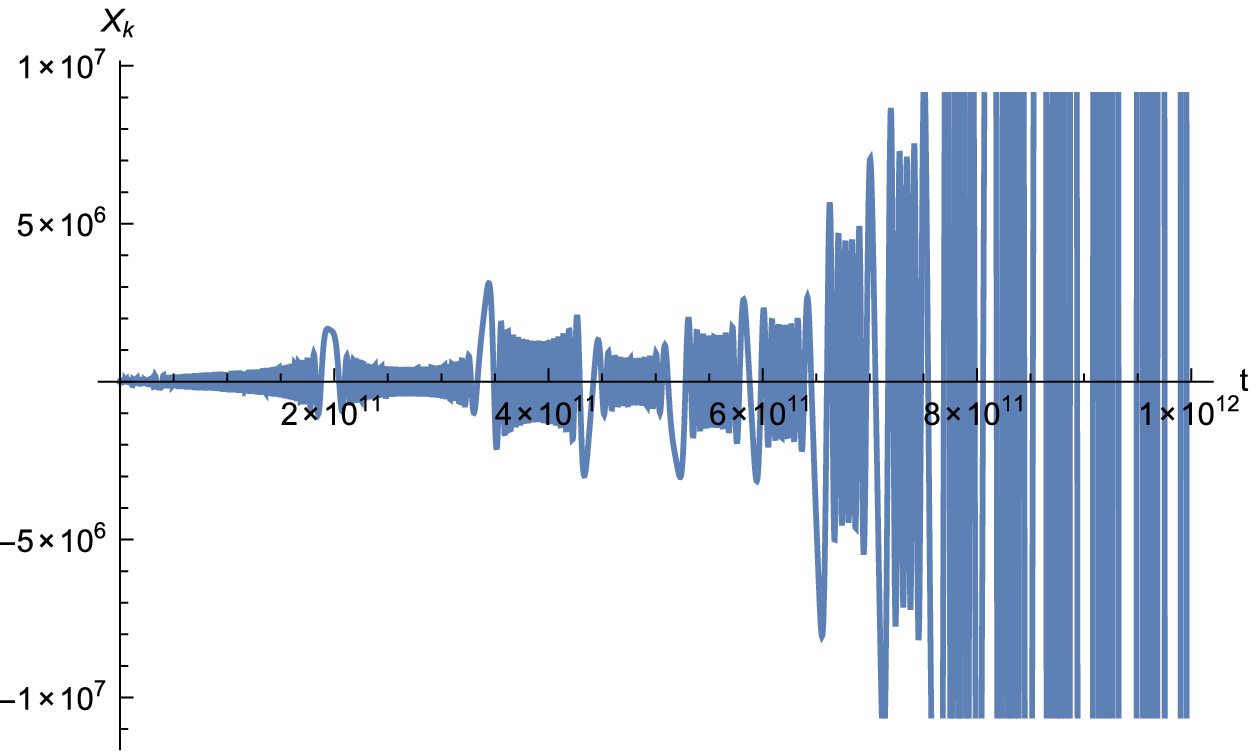}
\includegraphics[scale=0.5]{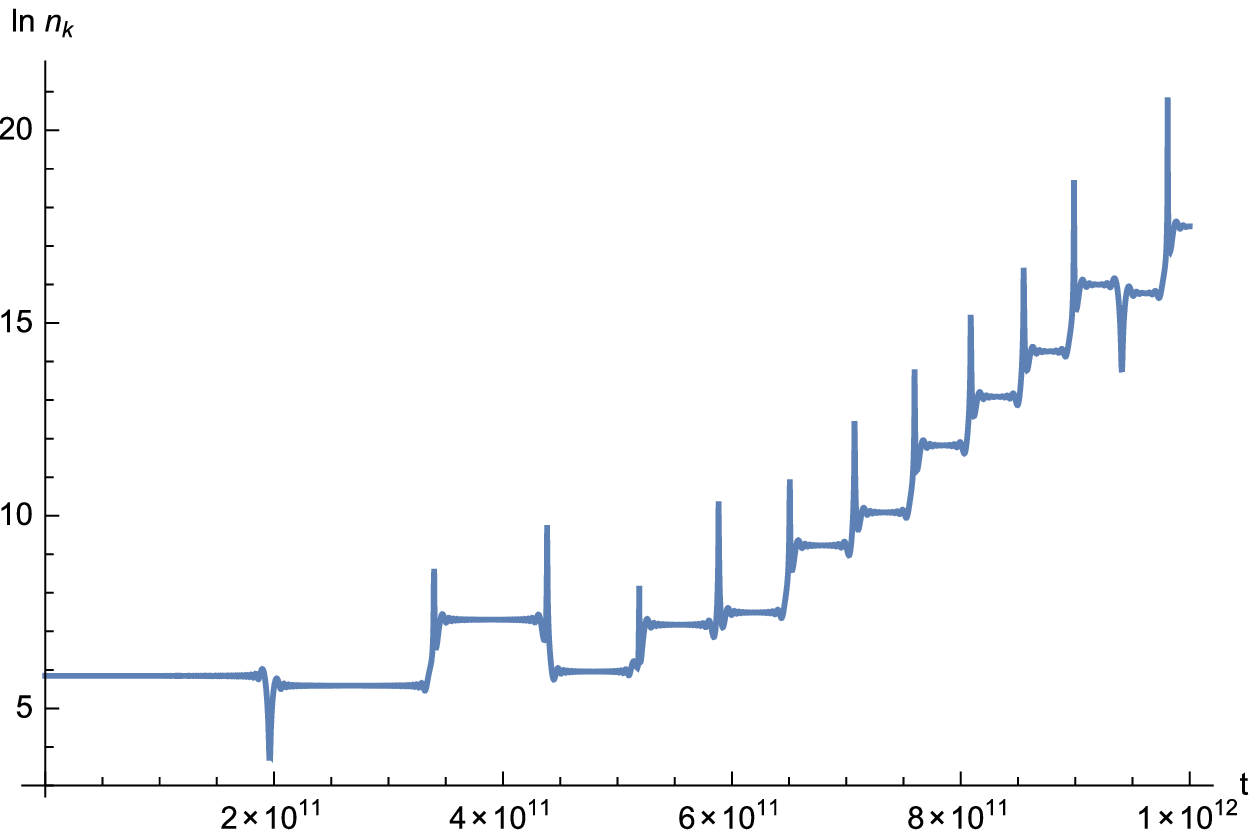}
\caption{The numerical plot of $X_k$ and $\ln n_k$ in terms of the cosmic time $t$ after curvaton domination. We choose the parameters as $\xi=1$, $M=10^{-4}M_p$, $m=10^{-6}M_p$, $g=4.8\times10^{-3}$. } \label{plot4}
\end{center}
\end{figure}

In Fig.s \ref{plot3} and \ref{plot4} we numerically calculated the creation of particles before and after curvaton domination, and plot the values of $X_k$ and $\ln n_k$ in terms of $t$. The numerical results are in good agreement with the theoretical analysis. In our plot we can see that in both cases $X_k$ oscillates rapidly due to the parametric resonance, and with the amplitude get higher and higher. The number density $n_k$ (in Logarithm) of both cases increase rapidly, though in some instances it decreases, due to the stochastic random process. Nevertheless, one can see that there are quantitative difference in the shapes as well as the increasing rates of $X_k$ and $n_k$ in the two cases, which is due to the different behavior of the source field $\varphi$ before and after curvaton domination.  

\begin{figure}[ht]
\begin{center}
\includegraphics[scale=0.7]{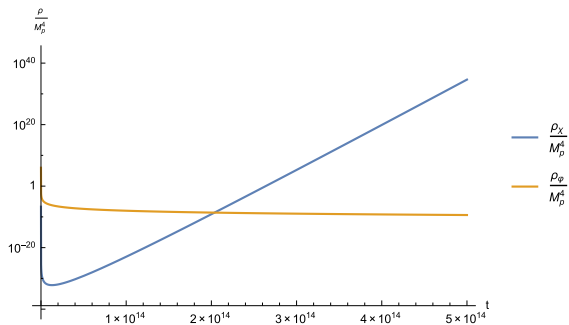}
\caption{The numerical plot of $\rho_\chi$ and $\rho_\varphi$ in terms of the cosmic time $t$. The time point when the two lines cross over corresponds to the time when the backreaction becomes important. We choose the parameters as $\xi=1$, $M=10^{-4}M_p$, $m=10^{-6}M_p$, $g=4.8\times10^{-3}$. } \label{backreaction}
\end{center}
\end{figure}

As a side remark, we roughly estimate the time when back-reaction of the produced particle $\chi$ becomes important, which can prevent further reheating process. To see this, one could define the energy density of the $\chi$ as:
\be
\rho_\chi\approx\langle n_\chi\rangle m_\chi~,
\ee
where $\langle n_\chi\rangle$ is calculated in (\ref{nchi}) while the effective mass of the $\chi$ particle is 
\be
m_\chi\approx\sqrt{\frac{g\Phi_0^2}{t^{3p-1}}\cos^2\left(\frac{mMt^2}{2\sqrt{6\xi}p}\right)-\frac{3}{2}\dot H-\frac{9}{4}H^2}
\ee
from Eq. (\ref{eomX}). The condition of back-reaction becoming important is $\rho_\chi\approx\rho_\varphi$, where $\rho_\varphi\approx m^2\varphi^2$. Since it is difficult to have an analytical solution of the time $t$, we perform the numerical calculation, and plot both $\rho_\chi$ and $\rho_\varphi$ (after neglecting the oscillation/resonance effect) in Fig. \ref{backreaction}. From the plot we can see that, the time when back-reaction becomes important is far after the curvaton domination.

\subsection{constraint on reheating temperature}
When reheating process completed, the curvaton will decay into the relativistic products, which scales as $T^4$ where $T$ is the temperature. Therefore the reheating temperature will be related to the final state of the curvaton field. Knowing the initial state of the curvaton field at where inflation has just ended, and the scaling of the curvaton field during reheating, one can estimate the reheating temperature and compare it to various data constraints. This procedure is initially done for normal curvaton mechanism given in \cite{Feng:2002nb}. Now we also use this method to estimate the reheating temperature in our model.

The curvaton may reheat the universe in two ways. Since we know from above that the energy density of curvaton scales as $a^0$ before domination, we have:
\be\label{rhophieq}
\rho_{\varphi}^{osc}\simeq\rho_{\varphi}^{eq}~
\ee
where the subscripts $osc$ and $eq$ denotes the value when $\varphi$ begins to oscillate and when it has the same amount of energy density with background, respectively. On the other hand, at the time when reheating completes we have:
\be\label{rhophirh}
\rho_{\varphi}^{rh}=\frac{g\pi^2}{30}T_{rh}^4~.
\ee
Therefore, if the curvaton reheats after its domination, then we have $\rho_{\varphi}^{eq}>\rho_{\varphi}^{rh}$, and the amount of difference depends on how long reheating will last after curvaton domination. From Eq.s (\ref{rhophieq}) and (\ref{rhophirh}) we have:
\be\label{Trh} 
T_{rh}<\left(\frac{30}{g\pi^2}\rho_{\varphi}^{osc}\right)^\frac{1}{4}~.
\ee

At the time when $\varphi$ ends slow-rolling and begins oscillating, the kinetic and potential term of $\varphi$ will be of the same order, namely $9\xi H^2\dot\varphi^2/M^2\sim m^2\varphi^2/2$, therefore
\be\label{rhophiosc}
\rho_{\varphi}^{osc}\simeq \frac{18\xi}{M^2}H^2\dot\varphi^2=18H_{osc}^2y_{osc}~,
\ee
where $y_{osc}$ is the value of $y (\equiv\xi\dot{\varphi}^2/M^2)$ at the oscillation time. So the reheating temperature is: 
\be\label{Trh2} 
T_{rh}<\left(\frac{540H_{osc}^2y_{osc}}{g\pi^{2}}\right)^{\frac{1}{4}}~.
\ee

The constraints on the Hubble parameter and $y$ can be obtained from the Planck constraints on scalar spectrum: $|{\cal P}_\zeta|\simeq2.2065^{+0.0763}_{-0.0738}\times10^{-9}$ (TT, TE, EE+lowP, $68\%$ CL) and the tensor/scalar ratio: $r\lesssim 0.10$ (TT, TE, EE+lowP, $95\%$ CL) \cite{Ade:2015lrj}. In our model, from \cite{Feng:2013pba} we have: 
\be
{\cal P}_\zeta\simeq\frac{3\sqrt{3}H_*^{2}\epsilon_*^2}{196\sqrt{7}\pi^{2}|y_*|}~,~~~{\cal P}_T\simeq\frac{2H_*^2}{\pi^2M_p^2}~.
\ee
for this case, where $``\ast"$ denotes the value evaluated at horizon crossing. These first give the constraints:
\be\label{constraintI}
\frac{\epsilon_*^2M_p^2}{|y_*|}\gtrsim 2\times10^3~,~~~H_*\lesssim 10^{-5}\pi M_p~,
\ee
from where one can see, if we set  $\epsilon_*\sim {\cal O}(10^{-2})$, we can get $y_*$ to be as low as $10^{-7}M_p^2$. As $y$ and $H$ varies slowly during inflation, one can roughly have $H_*\simeq H_{osc}$, $y_*\simeq y_{osc}$, therefore the reheating temperature can be set to be:
\bea\label{Trh3} 
T_{rh}&\lesssim&0.72\times10^{-4}\sqrt{\left(\frac{\epsilon}{0.01}\right)}M_p \nonumber\\
&\approx&0.879\times10^{15}\sqrt{\left(\frac{\epsilon}{0.01}\right)}\text{GeV}~,
\eea
which actually gives very loose constraint on the reheating temperature, comparing to that from the gravitino producing \cite{Kawasaki:1994af}. 

On the other hand, if curvaton decays and reheats before dominating the universe, one has $\rho_{\varphi}^{rh}\simeq\rho_{\varphi}^{osc}\simeq\rho_{\varphi}^{eq}$. From Eq.s (\ref{rhophirh}), (\ref{rhophiosc}), (\ref{Trh2}) we have:
\be 
T_{rh}\simeq\left(\frac{540H_{osc}^2y_{osc}}{g\pi^{2}}\right)^{\frac{1}{4}}~.
\ee
For perturbation in this case, we have:
\be
{\cal P}_\zeta\simeq\frac{3\sqrt{3}H_*^{2}\epsilon_*^2r^2}{784\sqrt{7}\pi^{2}|y_*|}~,~~~{\cal P}_T\simeq\frac{2H_*^2}{\pi^2M_p^2}~.
\ee
where $r\equiv\rho_\varphi/\rho_r$ is the ratio of energy densities of $\varphi$ field and the radiation. From Planck constraints, we have 
\be\label{constraintII}
\frac{\epsilon_*^2M_p^2}{|y_*|}\gtrsim \frac{8\times10^3}{r^2}~,~~~H_*\lesssim 10^{-5}\pi M_p~,
\ee
so we have 
\be 
T_{rh}\simeq0.623\times10^{15}\sqrt{\frac{\epsilon r}{0.01}}\text{GeV}~.
\ee
which gives a high reheating temperature. So if the constraints on gravitino are trustable, then this case will be ruled out \footnote{However, this constraint is based on the supersymmetry theory, which has not been proved yet.}. Note that recently, in Ref. \cite{Hardwick:2016whe}, the authors obtained a more stringent upper bound for reheating temperature of $T_{rh}<5\times10^{4}$GeV for minimal coupling curvaton models by Bayesian inference method \footnote{We thank the authors of Ref. \cite{Hardwick:2016whe} for pointing their paper to us.}.

Moreover, one can also constrain some of the parameters for the model, using that on Hubble parameter. From Eq. (\ref{meff}), one has $H_\ast\sim \sqrt{10mM/\sqrt{6\xi}}$ during inflation while $H_{osc}\simeq(m^2M^2/6\xi)^{1/4}$ when $\varphi$ begins to oscillate, which is given by $H_{osc}\simeq\bar{m}$. Since $H_\ast\simeq H_{osc}$, these two conditions actually do not differ much. From either (\ref{constraintI}) or (\ref{constraintII}), one can get the relation of $mM/\sqrt{6\xi}\lesssim 10^{-10}M_p$. That means, if we choose $m$ to be of $10^{-6}M_p$, we will get $M/\sqrt{6\xi}\lesssim 10^{-4}M_p$, which is consistent with our numerical calculations. 

\section{effects on curvature perturbations}
In this section, we discuss about how the resonance process could affect on the final curvature perturbations in our model, especially whether the points of $\dot\varphi=0$ will make the curvature perturbations diverge. This issue has been in debate in early works of canonical inflation reheating \cite{Finelli:1998bu, Lin:1999ps} (see also \cite{Jedamzik:2010dq, Easther:2010mr, Algan:2015xca} for further discussions), and has been revisited recently for NDC inflation reheating \cite{Myung:2016twf}. As we know, since there are more than one component in the universe, the perturbations generated by curvaton is basically isocurvature, which will transfer to curvature perturbation after curvaton domination. According to the original analysis on curvaton in \cite{Lyth:2001nq}, the curvature perturbation is defined as:
\be
\zeta=-H\frac{\delta\rho_{bg}+\delta\rho_\varphi}{\dot{\rho}_{bg}+\dot{\rho}_\varphi}
\ee
in spatial flat gauge. Assuming that the perturbations of the background field (inflaton) are negligible, we have $\delta\rho_{bg}\approx 0$. 

From continuity equations for both components, $\dot{\rho}=-3H(\rho+P)$. For background whose equation of state is assumed to be unity, we have $\dot{\rho}_{bg}=-6H\rho_{bg}$. Therefore for the time before curvaton dominating, one roughly have:
\be
\zeta=-H\frac{\delta\rho_\varphi}{\dot{\rho}_{bg}}=\frac{\rho_{\varphi,\varphi}\delta\varphi}{6\rho_{bg}}~.
\ee
Since $\rho_{bg}$ are non-zero, one can easily see that $\zeta$ will not diverge.

On the other hand, for the time after curvaton dominating, one have:
\be\label{zeta2}
\zeta=-H\frac{\delta\rho_\varphi}{\dot{\rho_\varphi}}=\frac{\rho_{\varphi,\varphi}\delta\varphi}{3H(\rho_\varphi+P_\varphi)}=-\frac{H}{\dot\varphi}\delta\varphi~.
\ee
where we have made use of Eqs. (\ref{rho}), (\ref{p}) as well as the relation:
\be
\rho_{\varphi,\varphi}=\frac{\dot \rho_{\varphi}}{\dot\varphi}=\frac{6\xi}{M^2}(2H^2\ddot\varphi+H\dot H\dot\varphi-3H^3\dot\varphi)~.
\ee 
So it seems that when $\dot\varphi$ passes through $0$ during oscillation, $\zeta$ might diverge. However, this is not true. From equation of motion for $\delta\varphi$ \cite{Feng:2013pba}:
\be
(az^2)^\cdot\dot{\delta\varphi}+az^2\ddot{\delta\varphi}-aQ\partial_i^2\delta\varphi+a^3m_{eff}^2\delta\varphi=0~,
\ee
where $z=a\sqrt{Q}/c_s\sim aH$, and $Q$, $c_s$ as well as $m_{eff}$ have all been defined in \cite{Feng:2013pba}. Making use of (\ref{zeta2}), we have:
\be\label{eomzeta}
(az^2)^\cdot(\frac{\dot\varphi}{H}\zeta)^\cdot+az^2(\frac{\dot\varphi}{H}\zeta)^{\cdot\cdot}-aQ\frac{\dot\varphi}{H}\partial_i^2\zeta+a^3m_{eff}^2\frac{\dot\varphi}{H}\zeta=0~.
\ee
On the other hand, from Eq. (\ref{eom}), one has:
\be
\ddot\varphi=-2\frac{\dot H}{H}\dot\varphi-3H\varphi-\frac{M^2V_{,\varphi}}{6\xi H^2}~
\ee
where $V_{,\varphi}=m^2\varphi$, and one more time derivative gives:
\bea
\dddot\varphi&=&-2\frac{\ddot H}{H}\dot\varphi-2\frac{\dot H}{H}\ddot\varphi+2\frac{\dot H^2}{H^2}\dot\varphi-3\dot H\dot\varphi-3H\ddot\varphi \nonumber\\ 
&&-\frac{M^2m^2}{6\xi H^2}\dot\varphi+\frac{M^2m^2\dot H}{3\xi H^3}\varphi~.
\eea

Expanding Eq. (\ref{eomzeta}) and making use of the above two equations, one finally have:
\be
\frac{a^3M^2m^2\varphi}{3\xi H}(\frac{2\dot H}{H}\zeta-\dot\zeta)=0~.
\ee
when $\dot\varphi=0$. It has the unique solution: $\dot\zeta=(2\dot H/H)\zeta$. However, since now the curvaton is dominant, we have $\dot H=-(\rho_\varphi+P_\varphi)/2=(\xi/M^2)(\dot H\dot\varphi^2+2H\dot\varphi\ddot\varphi-3H^2\dot\varphi^2)$, which is equal to zero when $\dot\varphi=0$. This can also be seen in Fig \ref{plot2}. So we still have $\dot\zeta=0$, and $\zeta$ will not diverge. Note that similar conclusion has been obtained in the early work of canonical single field inflation \cite{Lin:1999ps} \footnote{However, since the parameter sound speed squared $c_s^2$ contains $\epsilon\sim\dot H$ \cite{Feng:2013pba, Ema:2015oaa}, in the reheating era when $H$ is oscillating, $\dot H$ will also oscillate between positive and negative values, which causes negative $c_s^2$ and gradient instability for large-k modes during reheating \cite{Ema:2015oaa}. Since now we're interested the fluctuations outside horizon, the instability will not affect too much. We thank the anonymous referee for pointing this to us.}.

\begin{figure}[ht]
\begin{center}
\includegraphics[scale=0.5]{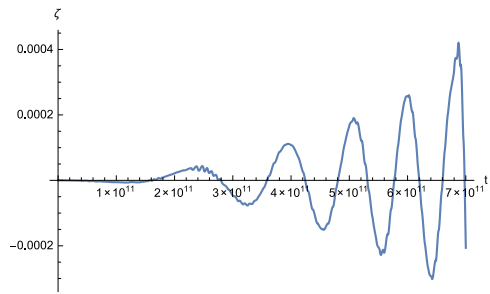}
\caption{The numerical plot of $\zeta$ in terms of the cosmic time $t$ before curvaton domination. We choose the parameters as $\xi=1$, $M=10^{-4}M_p$, $m=10^{-6}M_p$. The initial conditions for $\delta\varphi$ is $\delta\varphi_i=10^{-4}M_p$, $\dot{\delta\varphi}_i=0$.} \label{zetaplot1}
\end{center}
\end{figure}

\begin{figure}[ht]
\begin{center}
\includegraphics[scale=0.5]{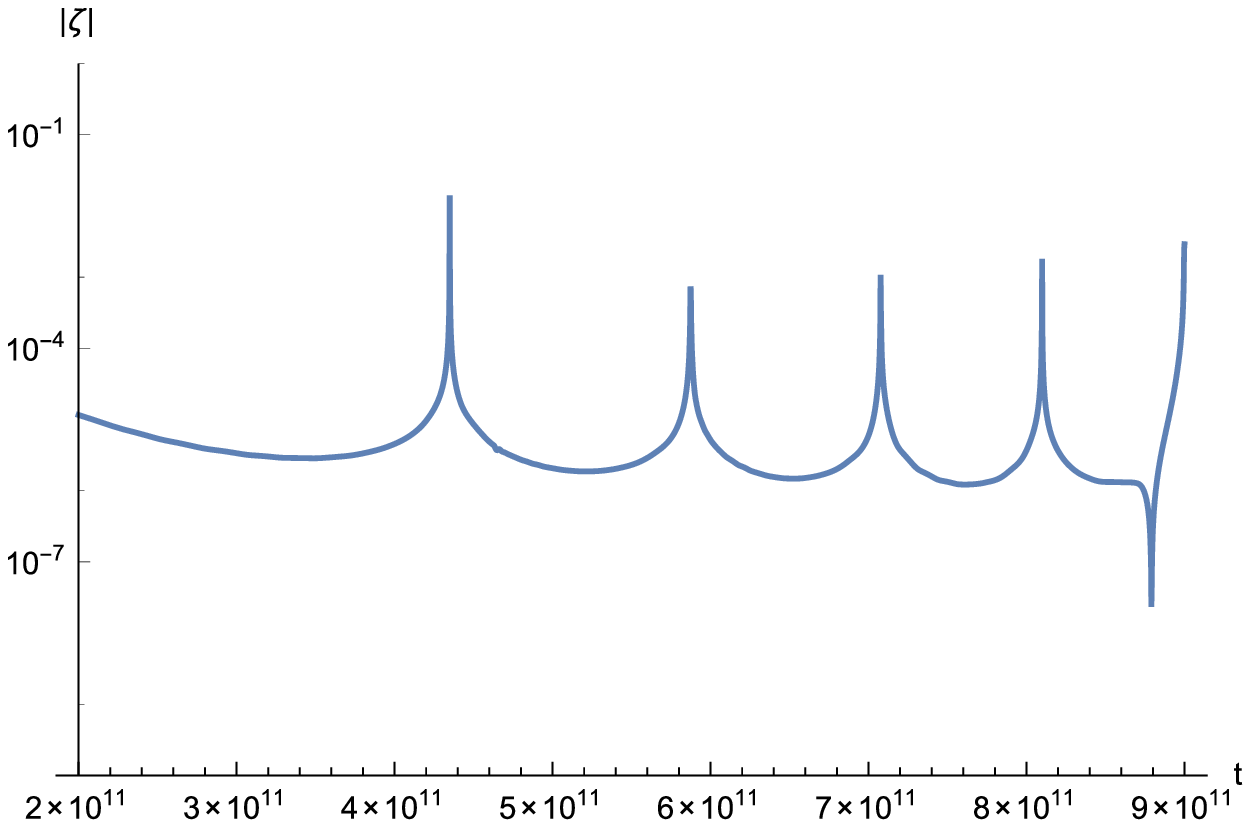}
\caption{The numerical plot of $|\zeta|$ in terms of the cosmic time $t$ after curvaton domination. We choose the parameters as $\xi=1$, $M=10^{-4}M_p$, $m=10^{-6}M_p$. The initial conditions for $\delta\varphi$ is $\delta\varphi_i=10^{-4}M_p$, $\dot{\delta\varphi}_i=0$.} \label{zetaplot2}
\end{center}
\end{figure}

\begin{figure}[ht]
\begin{center}
\includegraphics[scale=0.5]{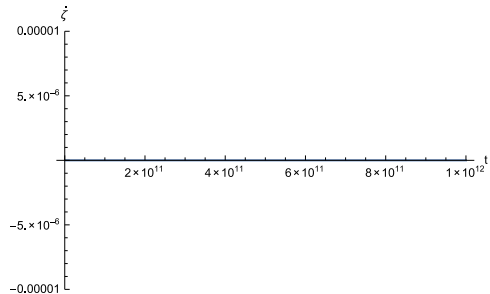}
\caption{The numerical plot of $\dot\zeta$ in terms of the cosmic time $t$ after curvaton domination. We choose the parameters as $\xi=1$, $M=10^{-4}M_p$, $m=10^{-6}M_p$. The initial conditions for $\delta\varphi$ is $\delta\varphi_i=10^{-4}M_p$, $\dot{\delta\varphi}_i=0$.} \label{dzetaplot2}
\end{center}
\end{figure}

We performed numerical calculations for both the two cases and plot the evolutions of $\zeta$ in Figs. \ref{zetaplot1} and \ref{zetaplot2}. One can see that, in Fig. \ref{zetaplot1}, $\zeta$ oscillates with an increasing amplitude. This can be explained as that, the perturbation generated before curvaton domination is of isocurvature type, which can be a source of $\zeta$. However in Fig. \ref{zetaplot2}, $\zeta$ behaves as a nearly constant. This is because when curvaton dominates and the isocurvature perturbation has been transformed into the curvature one, $\zeta$ will be a conserved quantity. Although there does have some features on $\zeta$, which may be due to the secondary effects of the resonance, numerical fluctuation or other unknown reasons, it will not be so bad to make it diverge. We also plot $\dot\zeta$ in Fig. \ref{dzetaplot2}, so one can see more clearly that, at the range of typical amplitude of $\zeta$ ($\sim 10^{-5}$), the variation of $\zeta$ can hardly be seen. 

\section{conclusions}     
In this paper, we investigated the reheating mechanism of the curvaton model nonminimally derivative coupled to gravity. We assume that the curvaton has a mass squared potential term, which can not only explain the observationally required tilt of the perturbation spectrum from scale invariance, but also provide a minimum around which the curvaton can oscillate. This gives an efficient way of particle creation, namely parametric resonance process. What is different from the reheating process of inflaton is that, the background evolutions before and after curvaton domination are not the same, and if the curvaton is nonminimally coupled, the averaged decaying rate of curvaton may also not be the same, which can affect the rate of particle creation. Though we didn't provide a rigid proof (which will be left for future work), this may be a general phenomenon.

For our model, as a specific case, we found that the averaged value of curvaton scales as a constant before curvaton domination, while proportional to time inverse after that. The particles, whose field interacts with the curvaton, will indeed be effectively created due to the resonance process. To confirm our analysis, we also numerically calculated the background evolution for the curvaton field and the increase of created particle numbers by the resonance.

Moreover, we also estimated the constraints on reheating temperature of our model, for both cases which the reheating completes before and after curvaton domination. We only got a upper limit on the temperature, which is looser than that given by the overproduction of gravitino. Finally, we investigated the curvature perturbation generated in our model. We showed that the curvature perturbation will be regular, although it is apparently divergent when $\dot\varphi$ goes to zero in each oscillation. Unlike that the amplitude of curvature perturbation is increasing before curvaton domination, which should be due to the source of isocurvature perturbation, after the curvaton domination it will remain an approximately constant. 

One can also investigate the back reaction of the created particle $\chi$ on the reheating process, which we leave for a future study. As an application, our work can be used either to compare cases of single and multiple field theories nonminimally coupled to gravity, or to compare cases of minimal and nonminimal couple theories in fixed degrees of freedom. 

\begin{acknowledgments}
We acknowledge Shinji Tsujikawa for suggestion to start this work and Yun-Song Piao, Yi-Fu Cai for useful discussions. The work of T.Q. is supported  in part by NSFC under Grant No: 11405069 and in part by by the Open Innovation Fund of Key Laboratory of Quark and Lepton Physics (MOE), Central China Normal University (No.:QLPL2014P01).
\end{acknowledgments}


\begin{thebibliography}{99}  
\bibitem{Amendola:1993uh}
  L.~Amendola,
  Phys.\ Lett.\ B {\bf 301}, 175 (1993)
  [gr-qc/9302010].
  
\bibitem{Deffayet:2009mn}
  C.~Deffayet, S.~Deser and G.~Esposito-Farese,
  Phys.\ Rev.\ D {\bf 80}, 064015 (2009)
  [arXiv:0906.1967 [gr-qc]];
  C.~Deffayet, X.~Gao, D.~A.~Steer and G.~Zahariade,
  Phys.\ Rev.\ D {\bf 84}, 064039 (2011)
  [arXiv:1103.3260 [hep-th]].

\bibitem{Capozziello:1999uwa}
  S.~Capozziello and G.~Lambiase,
  Gen.\ Rel.\ Grav.\  {\bf 31}, 1005 (1999)
  [gr-qc/9901051].

\bibitem{Cartier:2001is}
  C.~Cartier, J.~-c.~Hwang and E.~J.~Copeland,
  Phys.\ Rev.\ D {\bf 64}, 103504 (2001)
  [astro-ph/0106197].

\bibitem{Daniel:2007kk}
  S.~F.~Daniel and R.~R.~Caldwell,
  Class.\ Quant.\ Grav.\  {\bf 24}, 5573 (2007)
  [arXiv:0709.0009 [gr-qc]].

\bibitem{Sushkov:2009hk} 
  S.~V.~Sushkov,
  Phys.\ Rev.\ D {\bf 80}, 103505 (2009)
  [arXiv:0910.0980 [gr-qc]];
  E.~N.~Saridakis and S.~V.~Sushkov,
  Phys.\ Rev.\ D {\bf 81}, 083510 (2010)
  [arXiv:1002.3478 [gr-qc]].
  
\bibitem{Granda:2009fh}
  L.~N.~Granda,
  JCAP {\bf 1007}, 006 (2010)
  [arXiv:0911.3702 [hep-th]];
  L.~N.~Granda and W.~Cardona,
  JCAP {\bf 1007}, 021 (2010)
  [arXiv:1005.2716 [hep-th]];

\bibitem{Gao:2010vr}
  C.~Gao,
  JCAP {\bf 1006}, 023 (2010)
  [arXiv:1002.4035 [gr-qc]].
  
\bibitem{Germani:2010gm}
  C.~Germani and A.~Kehagias,
  Phys.\ Rev.\ Lett.\  {\bf 105}, 011302 (2010)
  [arXiv:1003.2635 [hep-ph]].

\bibitem{Chen:2010ru}
  S.~Chen and J.~Jing,
  Phys.\ Lett.\ B {\bf 691}, 254 (2010)
  [arXiv:1005.5601 [gr-qc]];
  S.~Chen and J.~Jing,
  Phys.\ Rev.\ D {\bf 82}, 084006 (2010)
  [arXiv:1007.2019 [gr-qc]].

\bibitem{Lin:2011zzd}
  K.~Lin, J.~Li and N.~Yang,
  Gen.\ Rel.\ Grav.\  {\bf 43}, 1889 (2011);
  J.~Li and Y.~Zhong,
  Int.\ J.\ Theor.\ Phys.\  {\bf 51}, 2585 (2012).

\bibitem{Banijamali:2012kq}
  A.~Banijamali and B.~Fazlpour,
  JCAP {\bf 1201}, 039 (2012)
  [arXiv:1201.1627 [gr-qc]].

\bibitem{Skugoreva:2013ooa} 
  M.~A.~Skugoreva, S.~V.~Sushkov and A.~V.~Toporensky,
  Phys.\ Rev.\ D {\bf 88}, 083539 (2013)
  Erratum: [Phys.\ Rev.\ D {\bf 88}, no. 10, 109906 (2013)]
  [arXiv:1306.5090 [gr-qc]].
  
\bibitem{Minamitsuji:2013ura} 
  M.~Minamitsuji,
  Phys.\ Rev.\ D {\bf 89}, 064017 (2014)
  [arXiv:1312.3759 [gr-qc]].

  
\bibitem{Myung:2015xha} 
  Y.~S.~Myung and T.~Moon,
  Int.\ J.\ Mod.\ Phys.\ D {\bf 24}, no. 14, 1550095 (2015)
  [arXiv:1502.03881 [gr-qc]];
  Y.~S.~Myung, T.~Moon and B.~H.~Lee,
  JCAP {\bf 1510}, no. 10, 007 (2015)
  [arXiv:1505.04027 [gr-qc]].
  
\bibitem{Yang:2015pga} 
  N.~Yang, Q.~Gao and Y.~Gong,
  arXiv:1504.05839 [gr-qc];
  N.~Yang, Q.~Gao and Y.~Gong,
  Int.\ J.\ Mod.\ Phys.\ A {\bf 30}, no. 28n29, 1545004 (2015);
  Y.~Zhu and Y.~Gong,
  Int.\ J.\ Mod.\ Phys.\ D {\bf 26}, 1750005 (2017)
  [arXiv:1512.05555 [gr-qc]].
  
\bibitem{Qiu:2015aha} 
  T.~Qiu,
  Phys.\ Rev.\ D {\bf 93}, no. 12, 123515 (2016)
  [arXiv:1512.02887 [hep-th]].
  
\bibitem{Cai:2016gjd} 
  Y.~Cai and Y.~S.~Piao,
  JHEP {\bf 1603}, 134 (2016)
  [arXiv:1601.07031 [hep-th]].
  
\bibitem{Feng:2013pba}
  K.~Feng, T.~Qiu and Y.~-S.~Piao,
  Phys.\ Lett.\ B {\bf 729}, 99 (2014)
  [arXiv:1307.7864 [hep-th]].
  
\bibitem{Feng:2014tka} 
  K.~Feng and T.~Qiu,
  Phys.\ Rev.\ D {\bf 90}, no. 12, 123508 (2014)
  [arXiv:1409.2949 [hep-th]].

\bibitem{Starobinsky:1980te} 
  A.~A.~Starobinsky,
  Phys.\ Lett.\  {\bf 91B}, 99 (1980).
  
\bibitem{Albrecht:1982mp} 
  A.~Albrecht, P.~J.~Steinhardt, M.~S.~Turner and F.~Wilczek,
  Phys.\ Rev.\ Lett.\  {\bf 48}, 1437 (1982).
  
\bibitem{Dolgov:1982th} 
  A.~D.~Dolgov and A.~D.~Linde,
  Phys.\ Lett.\ B {\bf 116}, 329 (1982).
  
\bibitem{Abbott:1982hn} 
  L.~F.~Abbott, E.~Farhi and M.~B.~Wise,
  Phys.\ Lett.\ B {\bf 117}, 29 (1982).
  
\bibitem{Kofman:1994rk} 
  L.~Kofman, A.~D.~Linde and A.~A.~Starobinsky,
  Phys.\ Rev.\ Lett.\  {\bf 73}, 3195 (1994)
  [hep-th/9405187];
  L.~Kofman, A.~D.~Linde and A.~A.~Starobinsky,
  Phys.\ Rev.\ D {\bf 56}, 3258 (1997)
  [hep-ph/9704452].
  
\bibitem{Traschen:1990sw} 
  J.~H.~Traschen and R.~H.~Brandenberger,
  Phys.\ Rev.\ D {\bf 42}, 2491 (1990);
  Y.~Shtanov, J.~H.~Traschen and R.~H.~Brandenberger,
  Phys.\ Rev.\ D {\bf 51}, 5438 (1995)
  [hep-ph/9407247].
  
\bibitem{Bassett:1997az} 
  B.~A.~Bassett and S.~Liberati,
  Phys.\ Rev.\ D {\bf 58}, 021302 (1998)
  Erratum: [Phys.\ Rev.\ D {\bf 60}, 049902 (1999)]
  [hep-ph/9709417];
  S.~Tsujikawa, K.~i.~Maeda and T.~Torii,
  Phys.\ Rev.\ D {\bf 60}, 063515 (1999)
  [hep-ph/9901306].
  
\bibitem{Feng:2002nb} 
  B.~Feng and M.~z.~Li,
  Phys.\ Lett.\ B {\bf 564}, 169 (2003)
  [hep-ph/0212213].
  
\bibitem{Dvali:2003em} 
  G.~Dvali, A.~Gruzinov and M.~Zaldarriaga,
  Phys.\ Rev.\ D {\bf 69}, 023505 (2004)
  [astro-ph/0303591];
  G.~Dvali, A.~Gruzinov and M.~Zaldarriaga,
  Phys.\ Rev.\ D {\bf 69}, 083505 (2004)
  [astro-ph/0305548].

\bibitem{Cai:2011ci} 
  Y.~F.~Cai, R.~Brandenberger and X.~Zhang,
  Phys.\ Lett.\ B {\bf 703}, 25 (2011)
  [arXiv:1105.4286 [hep-th]].
  
\bibitem{Sadjadi:2012zp} 
  H.~M.~Sadjadi and P.~Goodarzi,
  JCAP {\bf 1302}, 038 (2013)
  [arXiv:1203.1580 [gr-qc]];
  H.~M.~Sadjadi and P.~Goodarzi,
  JCAP {\bf 1307}, 039 (2013)
  [arXiv:1302.1177 [gr-qc]].
  
\bibitem{Ohashi:2012wf} 
  J.~Ohashi and S.~Tsujikawa,
  JCAP {\bf 1210}, 035 (2012)
  [arXiv:1207.4879 [gr-qc]].
  
\bibitem{Ghalee:2013ada} 
  A.~Ghalee,
  Phys.\ Lett.\ B {\bf 724}, 198 (2013)
  [arXiv:1303.0532 [astro-ph.CO]].
  
\bibitem{Jinno:2013fka} 
  R.~Jinno, K.~Mukaida and K.~Nakayama,
  JCAP {\bf 1401}, 031 (2014)
  [arXiv:1309.6756 [astro-ph.CO]].
    
\bibitem{Ema:2015oaa} 
  Y.~Ema, R.~Jinno, K.~Mukaida and K.~Nakayama,
  JCAP {\bf 1510}, no. 10, 020 (2015)
  [arXiv:1504.07119 [gr-qc]].
  
\bibitem{Myung:2016twf} 
  Y.~S.~Myung and T.~Moon,
  JCAP {\bf 1607}, no. 07, 014 (2016)
  [arXiv:1601.03148 [gr-qc]].
  
\bibitem{Bassett:2005xm} 
  B.~A.~Bassett, S.~Tsujikawa and D.~Wands,
  Rev.\ Mod.\ Phys.\  {\bf 78}, 537 (2006)
  [astro-ph/0507632].
  
\bibitem{Allahverdi:2010xz} 
  R.~Allahverdi, R.~Brandenberger, F.~Y.~Cyr-Racine and A.~Mazumdar,
  Ann.\ Rev.\ Nucl.\ Part.\ Sci.\  {\bf 60}, 27 (2010)
  [arXiv:1001.2600 [hep-th]].
  
\bibitem{Ade:2015lrj} 
  P.~A.~R.~Ade {\it et al.} [Planck Collaboration],
  arXiv:1502.02114 [astro-ph.CO].
  
\bibitem{Peebles:1998qn} 
  P.~J.~E.~Peebles and A.~Vilenkin,
  Phys.\ Rev.\ D {\bf 59}, 063505 (1999)
  [astro-ph/9810509].
  
\bibitem{Kawasaki:1994af} 
  M.~Kawasaki and T.~Moroi,
  Prog.\ Theor.\ Phys.\  {\bf 93}, 879 (1995)
  [hep-ph/9403364, hep-ph/9403061];
  R.~H.~Cyburt, J.~R.~Ellis, B.~D.~Fields and K.~A.~Olive,
  Phys.\ Rev.\ D {\bf 67}, 103521 (2003)
  [astro-ph/0211258];
  K.~Kohri, T.~Moroi and A.~Yotsuyanagi,
  Phys.\ Rev.\ D {\bf 73}, 123511 (2006)
  [hep-ph/0507245];
  M.~Kawasaki, F.~Takahashi and T.~T.~Yanagida,
  Phys.\ Rev.\ D {\bf 74}, 043519 (2006)
  [hep-ph/0605297].
  
\bibitem{Hardwick:2016whe} 
  R.~J.~Hardwick, V.~Vennin, K.~Koyama and D.~Wands,
  arXiv:1606.01223 [astro-ph.CO].
  
\bibitem{Finelli:1998bu} 
  F.~Finelli and R.~H.~Brandenberger,
  Phys.\ Rev.\ Lett.\  {\bf 82}, 1362 (1999)
  [hep-ph/9809490].
  
\bibitem{Lin:1999ps} 
  W.~B.~Lin, X.~H.~Meng and X.~M.~Zhang,
  Phys.\ Rev.\ D {\bf 61}, 121301 (2000)
  [hep-ph/9912510].
  
\bibitem{Jedamzik:2010dq} 
  K.~Jedamzik, M.~Lemoine and J.~Martin,
  JCAP {\bf 1009}, 034 (2010)
  [arXiv:1002.3039 [astro-ph.CO]].
  
\bibitem{Easther:2010mr} 
  R.~Easther, R.~Flauger and J.~B.~Gilmore,
  JCAP {\bf 1104}, 027 (2011)
  [arXiv:1003.3011 [astro-ph.CO]].
  
\bibitem{Algan:2015xca} 
  M.~T.~Algan, A.~Kaya and E.~S.~Kutluk,
  JCAP {\bf 1504}, no. 04, 015 (2015)
  [arXiv:1502.01726 [hep-th]].

\bibitem{Lyth:2001nq}
  D.~H.~Lyth and D.~Wands,
  Phys.\ Lett.\  B {\bf 524}, 5 (2002)
  [arXiv:hep-ph/0110002].

\end{thebibliography}
\end{document}